\documentclass[journal,comsoc]{IEEEtran}

\usepackage{amsmath,amsfonts}
\usepackage{algorithmic}
\usepackage{array}

\usepackage[caption=false,font=normalsize,labelfont=sf,textfont=sf]{subfig}

\usepackage{textcomp}
\usepackage{stfloats}
\usepackage{url}
\usepackage{verbatim}
\usepackage{graphicx}
\usepackage{xcolor}
\usepackage{cite}
\usepackage{balance}
\usepackage{booktabs}

\begin{document}
\bstctlcite{IEEEexample:BSTcontrol}

\title{Accelerating vRAN and O-RAN with SIMD: Architectural Perspectives and Performance Evaluation}

\author{Jaebum Park,~\IEEEmembership{Graduate Student Member,~IEEE},~Chan-Byoung Chae,~\IEEEmembership{Fellow,~IEEE}, and\\ Robert W. Heath Jr.,~\IEEEmembership{Fellow,~IEEE}

\thanks{J. Park and R. W. Heath, Jr.* are with the Department of Electrical and Computer Engineering, University of California, San Diego, La Jolla, CA 92093, USA (emails: \{jp261, rwheathjr\}@ucsd.edu).}
\thanks{C.-B. Chae* is with the School of Integrated Technology, Yonsei University, Seoul 03722, South Korea (email: cbchae@yonsei.ac.kr).} 
\thanks{*Co-corresponding authors: C.-B. Chae and R. W. Heath, Jr.}}

\maketitle

\begin{abstract} 
The evolution of radio access networks (RANs) toward virtualization and openness creates new opportunities for flexible, cost-effective, and high-performance deployments. Achieving real-time and energy-efficient baseband processing on commercial off-the-shelf platforms, however, remains a critical challenge. This article explores how single instruction multiple data (SIMD) architectures can accelerate RAN workloads. We first outline why key physical-layer functions, such as channel estimation and multiple-input multiple-output (MIMO) detection, are well aligned with SIMD’s data-level parallelism. We then present practical design guidelines and prototype results, showing
approximately 50\% improvement in throughput and more than a twofold improvement in energy efficiency compared to conventional CPU-only processing, while retaining programmability and ease of integration. Finally, we discuss open challenges in workload balancing and hardware heterogeneity, and highlight the role of SIMD as an enabling technology for flexible, efficient, and sustainable 6G-ready RANs.

\end{abstract}

\section{Introduction} 

\IEEEPARstart{T}{he} architecture of the radio access network (RAN) is continuing to evolve in the transition from 4G to 5G and beyond to 6G. This evolution is driven by several key factors. First, next-generation RANs must meet some performance requirements, including support for higher data rates and lower latency. Second, there is a strong need to reduce dependence on proprietary hardware through open architectures. Third, new deployment models like cloud RAN require more flexible and software-centric designs. These trends are motivating a shift away from hardware-defined solutions toward more modular and virtualized designs that can adapt to future requirements with greater agility.

\begin{figure*}[t]
    \centering
    \includegraphics[width=1.0\textwidth]{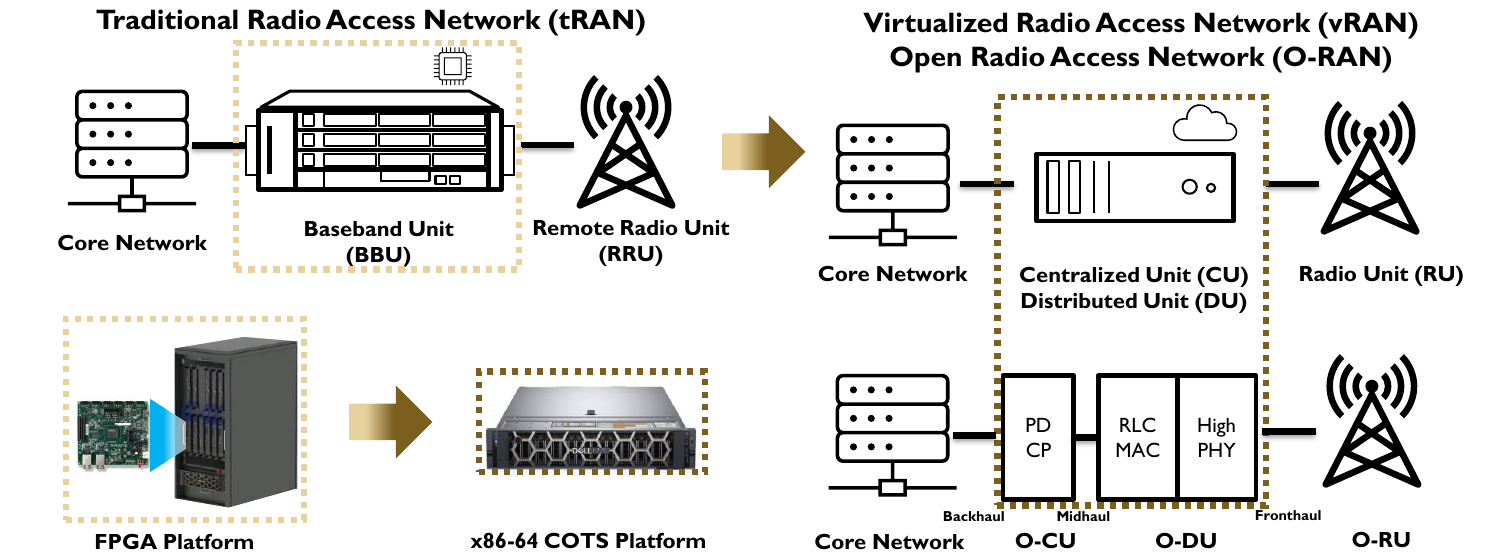}
    \caption{A general illustration of the transition from traditional RAN (tRAN) to virtualized and open RAN (vRAN/O-RAN) architectures. In tRAN, baseband processing is tightly coupled with proprietary hardware, limiting flexibility and scalability. In contrast, vRAN decouples hardware and software, enabling more agile deployments using commercial-off-the-shelf (COTS) servers. O-RAN further adopts a disaggregated architecture with standardized interfaces (e.g., O-CU, O-DU and O-RU). This shift also facilitates the integration of SIMD-based software acceleration, effectively addressing the computational requirements of distributed MIMO (D-MIMO) systems.}
    \label{fig:tRAN_vRAN}
\end{figure*}

In traditional RAN (tRAN) systems, baseband processing is largely implemented on dedicated hardware platforms such as field-programmable gate arrays (FPGAs), digital signal processors (DSPs), and application-specific integrated circuits (ASICs). While these platforms offer low latency and high efficiency, they are costly to develop, difficult to upgrade, and typically provided as tightly integrated vendor-specific solutions. Virtualized RAN (vRAN) addresses these limitations by implementing baseband functions in software running on commercial off-the-shelf (COTS) servers. Building on vRAN, the open RAN (O-RAN) framework adds standardized interfaces between RAN components, enabling multi-vendor interoperability and flexible system integration \cite{PoleseEtAlUnderstandingORANArchitecture2023}. Together, vRAN and O-RAN represent a fundamental shift toward software-defined RAN architectures, as illustrated in Fig.~\ref{fig:tRAN_vRAN}.

A key consequence of this shift is that vRAN and O-RAN rely primarily on general-purpose processors (GPPs), typically x86-64 CPUs, instead of dedicated baseband hardware. While GPUs have been explored for certain use cases, their higher power consumption and integration complexity have limited their widespread adoption in practical deployments. As a result, achieving real-time PHY-layer processing on CPU-based platforms requires efficient exploitation of CPU micro-architectural features. Single instruction, multiple data (SIMD) processing has therefore emerged as a critical enabler for modern vRAN and O-RAN baseband implementations, as it allows the same operation to be applied to multiple data elements in parallel. SIMD has been widely used in domains such as computer vision and image processing, where real-time parallel processing on thousands of pixels is required for applications including autonomous vehicles and robotics \cite{vanderMarkGavrilaRealtimeDenseStereo2006}. In wireless systems, SIMD can be applied to core baseband functions such as multiple-input multiple-output (MIMO) detection, which involves large vector and matrix computations \cite{singhSIMDenabledPhysicsinspiredMIMO2024}. This approach enables faster physical (PHY) layer processing without relying on external accelerators. Recent research has also addressed how to optimize data arrangement and memory access patterns in open-source vRAN platforms~\cite{WangHuEnablingEfficientSIMD2021}. In contrast, this paper targets algorithm-level acceleration and demonstrates the practical feasibility of SIMD-based MIMO detection under commercial 3GPP PHY parameters. The result highlights the potential for PHY-layer software optimization that reduces processing latency in real-world RAN deployments.

In this article, we introduce SIMD concepts to the wireless communications community and illustrate how they can be applied in modern RAN architectures. Although SIMD is a standard feature of general-purpose CPUs and widely used in other domains, it remains underutilized in many wireless PHY-layer implementations and is not addressed in current 3GPP or O-RAN specifications. We first review the evolution of RAN architectures with an emphasis on software-defined baseband processing. We then explain the fundamentals of SIMD and discuss why it is well suited for structured PHY-layer computations. Next, we present a design example of SIMD-accelerated MIMO detection under realistic 3GPP parameters and evaluate its potential in vRAN deployments. Finally, we discuss its effectiveness compared to GPUs and FPGAs and conclude with use cases and future research directions.

\section{The evolution of RAN architecture}

The architectural evolution toward vRAN and O-RAN shifts baseband processing from specialized hardware to software running on general-purpose CPUs. While this transition improves deployment flexibility, it also introduces significant computational challenges at the PHY layer, where strict real-time constraints must be satisfied using software-based implementations. As a result, efficiently exploiting CPU micro-architectural features becomes essential for sustaining real-time performance in software-defined RAN architectures. In this context, SIMD processing emerges as a practical and scalable acceleration technique that directly addresses these challenges. This section reviews the key architectural properties of vRAN and O-RAN that enable the practical deployment of SIMD-based acceleration for PHY-layer processing.

\subsection{The vRAN architecture} 

vRAN refers to the software-defined implementation of RAN functions that enables baseband processing to run on GPPs, such as x86-64 CPUs, on top of COTS computing platforms. By decoupling software from hardware, vRAN allows RAN workloads to operate on GPP-based servers and to be ported across processors with minimal effort through software recompilation. Industry developers can implement RAN functions using their familiar programming languages such as C or C++. In contrast, tRAN architectures perform PHY processing on hardwired fixed-function silicon, which requires hardware upgrades for new services. Legacy systems also depend on hard-coded hardware description languages (HDLs) such as Verilog. This approach ties the implementation design to specific hardware platforms and requires proprietary software tools, limiting innovation due to the need for developers with specialized expertise.

\begin{figure}[t]
  \centering
  \includegraphics[width=0.9\linewidth]{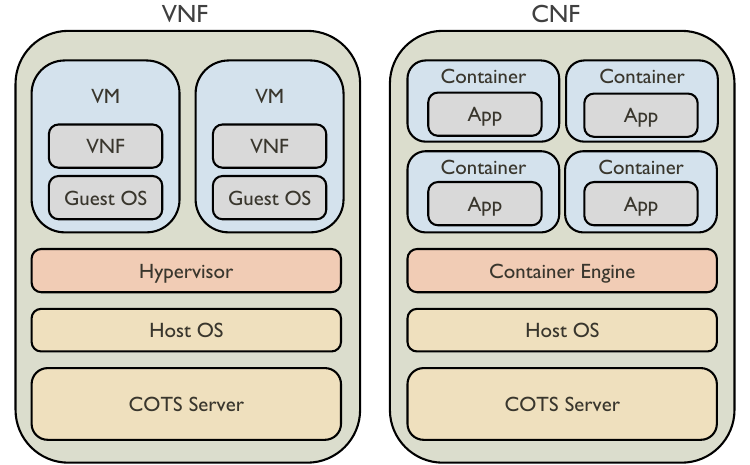}
  \caption{Architectural comparison between VNF- and CNF-based vRAN implementations. The VNF model relies on virtual machines with guest operating systems, while the CNF model employs lightweight containers orchestrated by a container engine.}
  \label{fig:vRAN}
\end{figure}

The evolution of vRAN architecture from virtualized network functions (VNFs) to cloud-native or containerized network functions (CNFs) significantly enhances operational efficiency. As illustrated in Fig.~\ref{fig:vRAN}, VNFs run within virtual machines (VMs) and require full guest operating systems (OSs) over a hypervisor, a software layer that creates and manages multiple VMs on a physical COTS server. While VNFs provided the initial step toward virtualization, they are resource-heavy and slower to scale. CNFs, in contrast, operate as lightweight containers generated by a container engine. This containerized structure eliminates the need for a full guest OS, enabling faster scaling, more efficient resource utilization, and improved software portability. CNF-based architectures also allow services to run across future processor generations through simple recompilation and lightweight container orchestration. Although we do not cover the CNF structure in detail, Kubernetes-based deployment is one of the most widely adopted container orchestrators for vRAN solutions. It further enables mobile network operators to manage and scale services more efficiently \cite{AttaouiEtAlVNFCNFPlacement2023}. This cloud-native evolution creates a flexible and portable software environment that supports SIMD-accelerated baseband processing on GPPs.

\subsection{The O-RAN architecture}

O-RAN extends the principles of vRAN by introducing open and standardized interfaces between functional RAN components, such as the radio unit (RU), distributed unit (DU), and centralized unit (CU) \cite{oranWG4CUS}. Although commercial adoption is still limited compared to tRAN and vRAN, O-RAN's main motivation is cost reduction through disaggregation of PHY-layer functions and increased vendor competition. In the long term, O-RAN adoption is expected to help maintain 5G leadership and influence the development of 6G systems.

At the functional level, O-RAN defines several functional split options that separate legacy RAN processing into the RU, DU, and CU, with standardized interfaces such as fronthaul and midhaul. These functional splits allow independent development and deployment of each unit, increasing flexibility in vendor selection and network design. Among the various functional split options, Option 7 represents a trade-off between fronthaul bandwidth requirements and RU complexity. Reducing the dimensionality of signals sent over the fronthaul is also an active research topic \cite{DemirEtAlCellFreeMassiveMIMO2024}, as it can significantly lower the processing load between the RU and DU.

From a PHY-layer perspective, O-RAN functional disaggregation places increasing computational responsibility on software-based DU and RU implementations. Meeting strict latency and throughput requirements under these constraints requires efficient utilization of internal CPU resources. In this context, SIMD-based acceleration is aligned with the software-centric DU/RU processing model in practical O-RAN deployments, supporting real-time performance.

\section{SIMD fundamentals and integration into PHY}

\subsection{SIMD fundamentals}

SIMD is a parallel computing architecture that enables a single instruction to operate on multiple data elements. It was originally introduced by Intel through extensions such as MMX, SSE, and AVX, and is now widely supported across modern CPU architectures. Compared to non-SIMD execution, SIMD enables fine-grained data-level parallelism directly on each CPU core and can achieve speedups of 8$\times$ to 16$\times$ for signal processing tasks.

\begin{figure*}[t]
  \centering
  \subfloat[]{\includegraphics[
  height=5cm, keepaspectratio]{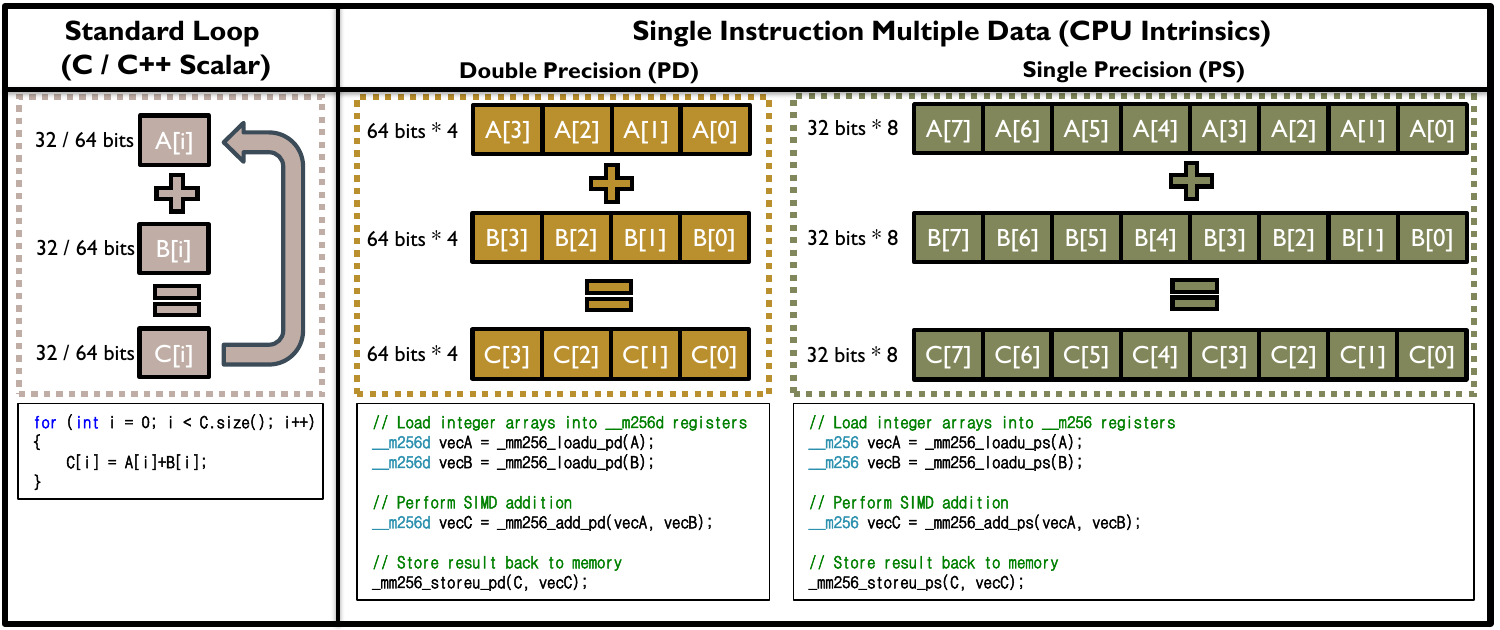} \label{fig:Scalar_SIMD}}
  \hfill
  \subfloat[]{\includegraphics[width=8.3cm,
  height=5cm,
  keepaspectratio]{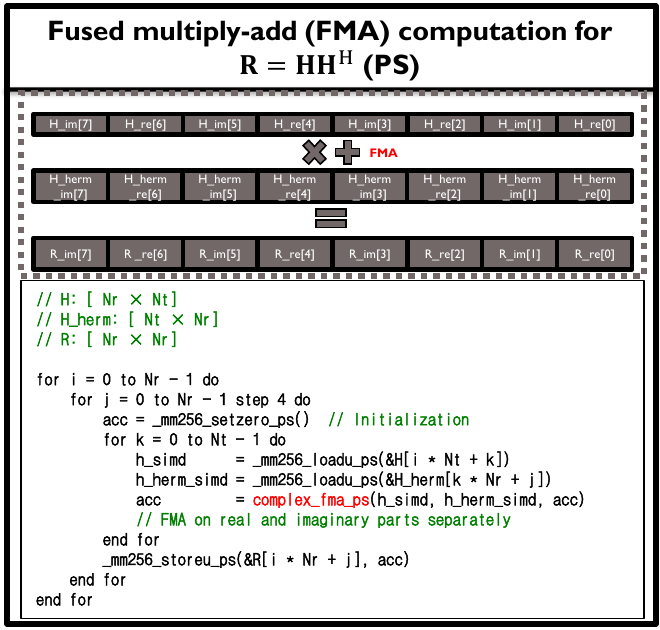} 
  \label{fig:FMA_SIMD}}
  \caption{Illustration of SIMD fundamentals. (a) Element-wise array addition, comparing standard scalar loop with SIMD intrinsics in both single- and double-precision to demonstrate parallelism. (b) Fused multiply‑add (FMA) computation for efficiently calculating the covariance matrix using vectorized operations.}
  \label{fig:SIMD_Tutorial}
\end{figure*}

SIMD acceleration is realized by defining specialized data types (e.g., $\texttt{\_\_m256}$ and $\texttt{\_\_m128}$ for Intel SIMD extensions) and applying vectorized arithmetic instructions such as \texttt{add}, \texttt{subs}, \texttt{mul}, and \texttt{div}. Fig.~\ref{fig:SIMD_Tutorial}\subref{fig:Scalar_SIMD} demonstrates this concept using element-wise array addition. The left side shows a scalar loop, which processes one element at a time. The right side shows SIMD intrinsics (e.g., $\texttt{\_\_m256\_add\_pd}$ for double precision and $\texttt{\_\_m256\_add\_ps}$ for single precision) that process multiple elements in parallel in one CPU instruction like four packed-double (PD) or eight packed-single (PS) values in a 256-bit register. This example highlights how SIMD exploits data-level parallelism to accelerate computation. A key practical advantage of SIMD intrinsics is their compatibility with standard C/C++ environments, including compilers (e.g., GCC) and debuggers (e.g., GDB). This compatibility simplifies development and debugging across Intel CPU generations and similar architectures. Provided as C function-like application programming interfaces (APIs), SIMD intrinsics directly map to specific assembly instructions and offer better readability than raw assembly. These features make SIMD a portable and sustainable option for PHY-layer processing.

The performance gains achieved by SIMD depend strongly on how efficiently algorithms are optimized for vectorized execution. Modern CPUs support extended SIMD instruction sets such as Intel AVX2 and AVX-512, which provide 256-bit and 512-bit vector registers, respectively. These extensions enable high-throughput floating-point operations (FLOPs), essential for computationally intensive MIMO workloads involving dense matrix and vector algebra. From an implementation perspective, fully leveraging SIMD requires careful memory alignment, loop unrolling, and efficient load/store operations to maximize the benefit of register-level bandwidth. Moreover, SIMD allows precise control over the trade-off between performance and numerical accuracy. For example, packing single-precision (32-bit) floating-point elements instead of double-precision (64-bit) floating-point elements into a 256-bit register $\texttt{\_\_m256}$ can nearly double the throughput. In many PHY-layer signal processing tasks, 16-bit precision per real and imaginary component is sufficient to represent signal values, resulting in minimal degradation in detection performance. At the same time, SIMD acceleration introduces non-trivial engineering challenges. Vector widths do not always align with algorithmic dimensions and require tail-handling for residual elements. If not handled properly, a subset of samples may be processed incorrectly, potentially degrading performance in terms of block error rate (BLER).

SIMD acceleration can improve energy efficiency primarily by reducing execution time. A recent system-level study shows that significant energy savings can be achieved through CPU-centric optimization under strict real-time constraints, without relying on GPU offloading~\cite{KaliaEtAlNSDI2025vRAN}. In addition, an empirical benchmarking study demonstrates that although SIMD-optimized workloads may exhibit slightly higher instantaneous power consumption, a 2--2.3$\times$ speedup can translate into nearly 50\% overall energy savings on both Intel x86 and ARM platforms~\cite{CriadoEtAlCoreNEURONPerformanceEnergy2020}. These results suggest that SIMD can serve as an energy-efficient acceleration mechanism for computation-heavy PHY-layer tasks in CPU-centric vRAN architectures.

\subsection{SIMD integration into PHY}

SIMD shows its benefits when applied to heavy PHY operations in modern RAN systems. In the context of vRAN and O-RAN, SIMD can accelerate a wide range of PHY-layer blocks, including MIMO detection, precoding and channel estimation. Among these, MIMO detection is chosen as the primary focus of this article because it involves repetitive complex-valued operations applied across multiple receive antennas and spatial streams, making it both computationally demanding and well suited for SIMD-based acceleration. By enabling parallel processing of multiple antenna-domain samples using a single instruction, SIMD directly exploits the inherent data-level parallelism of MIMO signal processing. As shown in Fig.~\ref{fig:SIMD_Tutorial}\subref{fig:FMA_SIMD}, the fused multiply-add (FMA) example maps directly to intermediate operations in MIMO receivers such as covariance calculation. These repetitive linear algebraic tasks benefit significantly from SIMD's data-level parallelism through vectorized additive and multiplicative instructions. We use MIMO detection as a case study to demonstrate how SIMD can be systematically integrated into broader PHY-layer pipelines in vRAN and O-RAN systems. Offloading these tasks to SIMD-enabled software modules allows vRAN systems to achieve lower latency and higher throughput without the need for GPUs or FPGAs.

\begin{figure}[t]
  \centering
  \includegraphics[width=\linewidth]{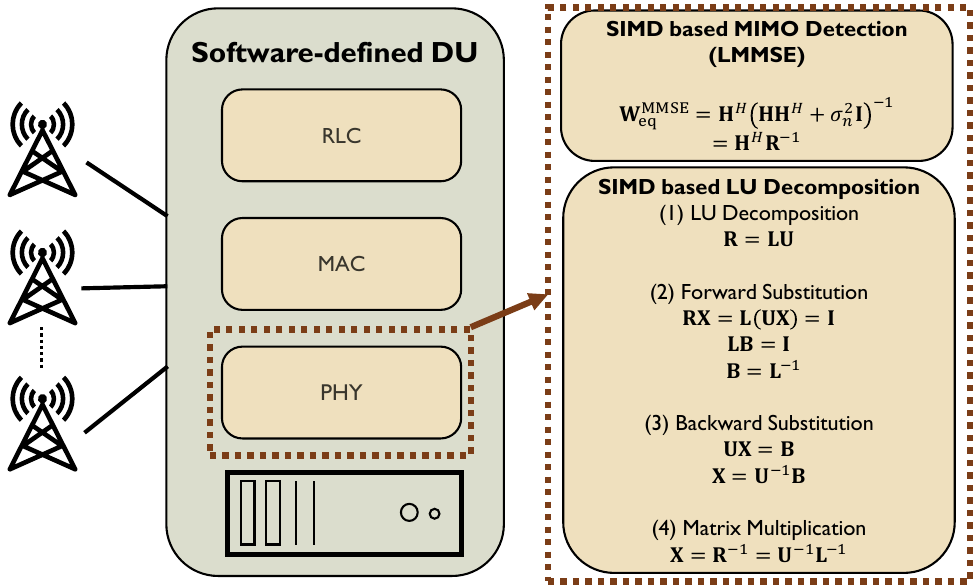}
  \caption{SIMD-accelerated MIMO detection algorithm implemented in the software-defined DU. The LMMSE detection process, consisting of LU decomposition, forward/backward substitution, and matrix inversion, is efficiently executed using SIMD instructions by exploiting vector-level parallelism.}
  \label{fig:MIMO_SIMD}
\end{figure}

SIMD-based baseband processing also runs directly on CPU cores, minimizing PCIe-related data transfer overhead and enabling tight timing control. These benefits are especially valuable in commercial vRAN and O-RAN systems that rely on fully software-defined signal chains running on COTS servers. A recent white paper demonstrates the viability of this approach by implementing a SIMD-accelerated polar decoder using AVX-512 for 5G control channels, showing that even latency-sensitive L1 modules can be handled
entirely in software on x86-64 platforms \cite{intelPolarWhitepaper2021}. It is worth noting that this prior work focuses on relatively simple decoding tasks with low computational complexity and small payload sizes, such as control channels, which align well with SIMD execution.

We extend this approach to large-scale data processing, where MIMO detection involves far larger payloads and tighter transmission time interval (TTI) constraints. By demonstrating that SIMD can efficiently handle these demanding baseband operations, we highlight the broader applicability of software-defined PHY acceleration in real-time, high-throughput vRAN systems. In this context, CPU-integrated SIMD acceleration is particularly well suited for indoor vRAN deployments and small-cell base stations, where the use of FPGA- or ASIC-based accelerators is often less attractive due to cost, power, and form-factor constraints. For such scenarios, optimizing PHY-layer processing using SIMD on general-purpose CPUs provides a practical and cost-effective solution.

\section{Performance evaluation of SIMD-accelerated MIMO detection}

\begin{figure*}[t]
  \centering
  \subfloat[]{\includegraphics[width=0.26\textwidth]{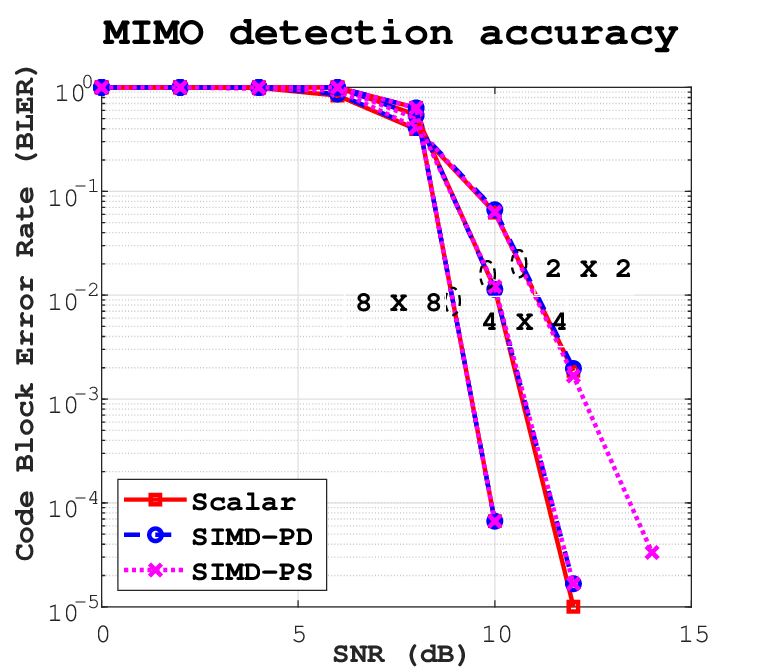} \label{fig:SIMD_Performance}}
  \hfill
  \subfloat[]{\includegraphics[width=0.72\textwidth]{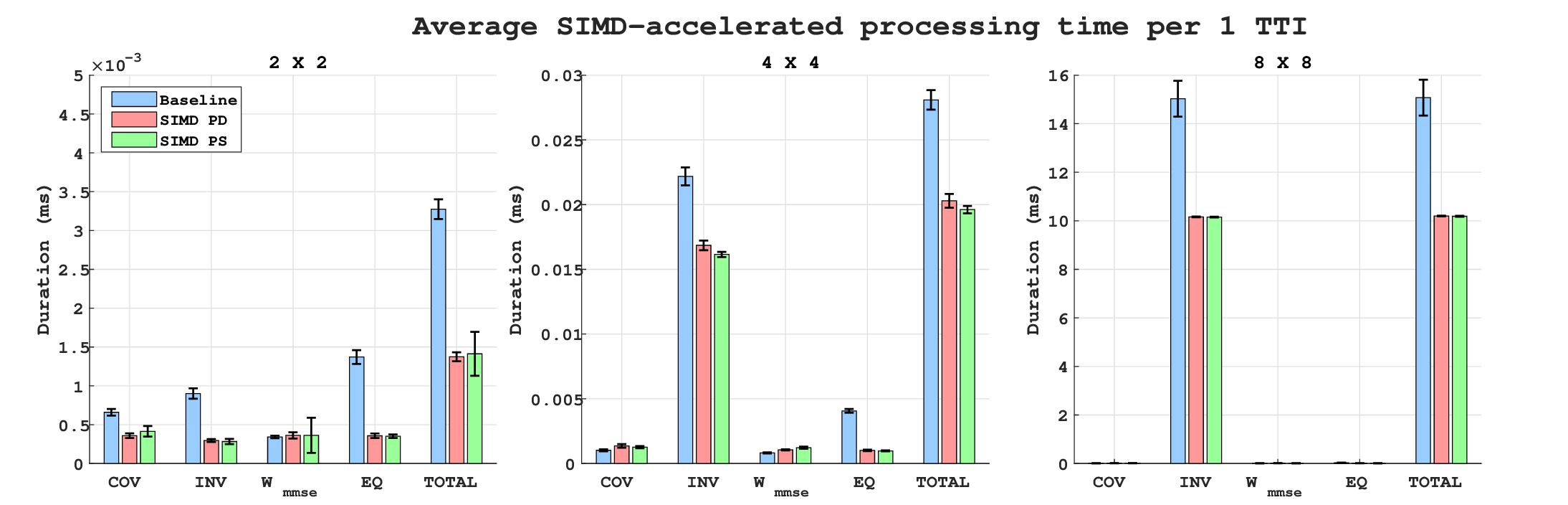} \label{fig:Acceleration}}
  \caption{Performance analysis of SIMD-based LMMSE MIMO detection. (a) Accuracy comparison between packed-single (PS) and packed-double (PD) modes under 3GPP NR and TDL-C channels. (b) Processing time breakdown across different MIMO configurations, showing that matrix inversion dominates runtime and benefits most from SIMD acceleration.}
  \label{fig:SIMD_Combined}
\end{figure*}

To evaluate the effectiveness of SIMD acceleration in realistic MIMO scenarios, we developed a link-level simulator configured with 3GPP NR parameters. This C++-based simulator emulates the behavior of commercial base station equipment running on COTS servers. It includes key PHY modules compliant with NR standards and serves as a reference platform to evaluate the performance gains of SIMD acceleration under conditions representative of current and near-future vRAN and O-RAN deployments.

Figure~\ref{fig:MIMO_SIMD} depicts the detailed structure of the linear minimum mean square error (LMMSE) MIMO detection algorithm. It highlights that matrix inversion remains one of the most computationally demanding tasks. To reduce this burden, we implement LU decomposition, which factorizes matrices into lower (L) and upper (U) triangular components \cite{GolubVanLoanMatrixComputations2013}. This transformation simplifies the inversion problem into two sequential steps: forward substitution and backward substitution, consisting of simple arithmetic operations that are highly suitable for SIMD vectorization. While modern CPUs and GPUs can optimize LU decomposition using tuned numerical libraries, such acceleration typically targets generic matrix operations. By explicitly vectorizing each computation step with SIMD intrinsics, user-managed implementations better exploit data-level parallelism in fixed-dimension PHY workloads. This enables additional performance gains beyond scalar execution while preserving deterministic behavior.

In the simulation, we configure LDPC channel encoding/decoding and MIMO-OFDM transmission using 15 kHz subcarrier spacing and 60 resource blocks (RB), corresponding to 720 subcarriers. We use modulation and coding scheme (MCS) index 9, as specified in \cite{3gpp38214}, to measure the average processing time per TTI for MIMO detection over 10,000 TTIs. The channel model is based on the 3GPP TDL-C specification. We evaluate signal detection performance using an LMMSE receiver, which involves inverting the received signal covariance matrix, $\mathbf{R}$. Fig.~\ref{fig:SIMD_Combined}\subref{fig:SIMD_Performance} illustrates that SIMD PS instructions achieve detection accuracy nearly identical to that of PD, despite the reduced arithmetic precision. This observation confirms that in practical MIMO scenarios, significant acceleration can be obtained without compromising detection performance. As the dimension of $\mathbf{R}$ increases with antenna scaling, we assess the computational efficiency of SIMD optimization in handling the resulting matrix operations. The evaluation was conducted on an Intel Core i9-14900KF CPU with AVX2 support. This configuration enables 256-bit SIMD operations using intrinsic data types such as $\texttt{\_\_m256}$. These instructions allow parallel processing of multiple complex numbers per clock cycle, up to four 64-bit complex numbers (with 32-bit real and imaginary parts each) or eight 32-bit complex numbers (with 16-bit real and imaginary parts each).

As shown in Fig.~\ref{fig:SIMD_Combined}\subref{fig:Acceleration}, SIMD-based MIMO processing demonstrates approximately 50\% speedup compared to a baseline vRAN configuration without SIMD acceleration, primarily due to optimization of the matrix inversion.
This performance gain is observed across different MIMO sizes ($2\times2$, $4\times4$, and $8\times8$), where matrix inversion accounts for the largest portion of the processing time and benefits the most from SIMD acceleration.
For example, in the $2\times2$ MIMO configuration, the detection processing time is reduced from approximately $3.3\,\mu\text{s}$ to less than $1.5\,\mu\text{s}$ when SIMD acceleration is enabled.
We profiled the acceleration gains for each sub-module of LMMSE detection block rather than isolated vector processing alone. As a result, the observed speedups are lower than the theoretically ideal SIMD speedups, since they include data movement, load/store latency, and interface overheads within the detection block. Nevertheless, a consistent speedup of more than twofold in the computationally dominant stage reduces overall L1 processing latency. Since linear MIMO detection is the primary computational bottleneck in L1 processing, SIMD acceleration offers the greatest opportunity for reducing end-to-end processing latency. When reducing data precision from double to single, the system maintains comparable accuracy while achieving additional acceleration gains. As vRAN manufacturers increasingly adopt high-performance CPUs, using AVX-512 instructions with 512-bit vector registers can further maximize the processing speed. This trend highlights the potential of lightweight, software-native techniques for meeting real-time PHY processing deadlines in vRAN deployments.

We also analyze the peak PHY-layer and evaluate the scalability of SIMD acceleration under different MIMO configurations. To estimate the theoretical peak rates, we consider a 5G NR system operating with MCS index 27, which uses 256-QAM modulation and a high coding rate. Leveraging the MCS, RB allocation, and the number of data streams, we can calculate the total PHY-layer payload bits over 1 TTI, known as transport block size (TBS) \cite{3gpp38214}.  With a 60 RB allocation and 15 kHz subcarrier spacing, each $2\times2$ MIMO carries approximately 139,376 bits per slot (or 139.4 kbit over 1 ms), resulting in a peak physical-layer throughput of 139.4 Mbps. Extending this to a $4\times4$ MIMO configuration yields about 278,776 bits per slot, or 279 Mbps. These values represent realistic PHY-layer payloads under 1 ms TTI constraints.

Targeting $4\times4$ MIMO for indoor deployments, the above configuration is appropriate, and our vRAN prototype demonstrates that SIMD acceleration reduces PHY processing time by more than 50 \% compared to a scalar implementation. This enables one of the most demanding tasks, such as $4\times4$ MIMO detection, to be completed within 0.03 ms. The processing time accounts for only about 3\% of the 1 ms TTI, leaving a substantial margin for other baseband operations. Such performance confirms that all PHY computations can be completed within the 1 TTI deadline. The evaluation results validate that software-defined base stations can meet real-time constraints without external accelerators. This efficiency reduces the execution time of intensive matrix operations and can therefore translate into significant overall energy savings at the system level. Higher MIMO configurations and wider bandwidths can be supported on modern CPUs with wider SIMD capabilities, suggesting that the proposed approach can scale beyond the evaluated configuration.

\section{High-level design trade-offs}

\begin{table*}[t]
    \centering
\caption{Comparison of PHY acceleration techniques for vRAN}
\label{tab:accel_compare}
\begin{tabular}{lccc}
\toprule
\textbf{Metric} & \textbf{SIMD (CPU)} & \textbf{SIMT (GPU)} & \textbf{FPGA} \\
\midrule
Latency & Low & Medium & Very Low \\
Flexibility & High & Medium & Low \\
Deployment Effort & Low & High & Very High \\
Energy Efficiency & Medium & Low & High \\
Scalability & High & High & Low \\
Toolchain / Maintenance & Standard (C/C++) & CUDA/OpenCL & Vendor-specific \\
\bottomrule
\end{tabular}
\end{table*}

In vRAN systems, several acceleration options exist for PHY-layer processing, including CPU-based SIMD, GPU-based single instruction, multiple threads (SIMT), and FPGA solutions. Among these, SIMD operates directly on general-purpose CPUs and is well-suited for latency-sensitive workloads. It also executes vectorized instructions on CPU registers with minimal control overhead and data transfer latency.

Table \ref{tab:accel_compare} demonstrates the strengths and weaknesses of GPUs in vRAN PHY acceleration. GPUs offer massive thread-level parallelism and can achieve very high throughput. In signal processing tasks, however, they often require substantial software tuning to reach peak performance. SIMT execution also faces warp divergence, where threads within a warp take different control paths, and scheduling overhead. These factors introduce latency variability and reduce energy efficiency. They also make GPUs less suitable for latency-critical real-time workloads.

FPGA-based solutions provide deterministic latency and excellent energy efficiency. They also deliver high timing precision, which is essential for time-critical PHY-layer tasks. These benefits come with trade-offs, however, as they require long development cycles, reliance on proprietary toolchains, and limited flexibility after deployment. Such constraints reduce agility and slow adaptation to evolving RAN requirements, particularly in software-defined environments.

SIMD applies uniform vector operations across aligned data vectors. Unlike ASIC-based accelerators such as tensor processing units (TPUs) or neural processing units (NPUs), which are optimized for batch-oriented AI inference, SIMD supports continuous and low-latency stream processing tightly coupled with the CPU. MIMO PHY-layer algorithms operate on real-time data streams and must adapt dynamically to channel conditions and scheduling decisions, which differ from batch-based AI workloads. By avoiding the data-movement overhead of external accelerators, SIMD enables real-time processing with low latency while maintaining programmability and flexibility. Overall, SIMD presents a balanced combination of scalability, low integration overhead, and compatibility with standard x86-64 server platforms. This balance makes it attractive for near-term 5G and emerging 6G RAN deployments, where ease of deployment, real-time performance, and software sustainability are crucial.

\section{Future outlook}

SIMD-based acceleration represents a software-native, CPU-integrated mechanism that aligns with vRAN and O-RAN architectures envisioned for 6G. Importantly, modern CPUs already incorporate advanced SIMD instruction sets, making these acceleration capabilities immediately available on existing COTS platforms without requiring new processor designs. By enabling efficient and deterministic PHY-layer processing on COTS platforms, SIMD complements future heterogeneous RAN designs without relying on external accelerators. Using an NR-compliant end-to-end link-level simulator, this work demonstrated that SIMD can significantly accelerate key PHY-layer functions such as matrix inversion and signal detection on COTS platforms. These results highlight SIMD as a practical enabler for software-defined RAN systems, thereby motivating future studies on its application to broader use cases.

\subsection{Use cases and future research directions}

The application of SIMD goes beyond PHY-layer testing and extends to diverse deployment scenarios. Several representative examples include: 

\begin{itemize}
    \item \textbf{Distributed MIMO (D-MIMO)}: In D-MIMO~\cite{RodriguezSanchezEtAlDecentralizedMassiveMIMO2020}, spatially distributed RUs cooperate to serve users, creating high fronthaul traffic and a heavy computational load on the DU. SIMD can accelerate fronthaul compression using block floating-point (BFP) quantization and baseband processing. These support real-time coordination across DUs and enable efficient, scalable D-MIMO operation.

    \item \textbf{Green 6G RAN}: As 6G targets sustainability and carbon neutrality, energy efficiency in RAN architectures is a critical concern~\cite{ericsson6gEnergy2024}. SIMD-based processing addresses this by executing PHY-layer tasks efficiently on GPPs. It reduces data movement and avoids device passthrough overhead. Its deterministic execution and tight software integration make it well-suited for energy-aware baseband pipelines in Green RAN deployments.

    \item \textbf{Artificial intelligence (AI)-RAN}: Although SIMD acceleration operates at the PHY implementation level, key performance metrics such as processing load, latency, and CPU utilization can be abstracted at the O-DU and reported to the Near-RT RIC via standardized E2 interfaces using existing service models (e.g., E2SM-KPM)  \cite{BalasubramanianEtAlRICRANIntelligent2021}. For example, increases in CPU utilization or latency reported through E2SM-KPM can trigger Near-RT RIC xApps to adjust DU resource allocation or scheduling policies to maintain real-time PHY performance.
\end{itemize}

\subsection{The road ahead: hybrid and heterogeneous computing}

Hybrid and heterogeneous computing techniques will play a critical role in future RAN architectures. This is particularly for large-scale deployments with massive antenna arrays anticipated in 6G, where memory bandwidth and data movement increasingly constrain scalability. In such regimes, SIMD should be viewed as one component of a broader acceleration strategy. This strategy combines (1) SIMD for deterministic and fine-grained PHY acceleration at the CPU level, (2) FPGAs for ultra-low-latency or high-throughput tasks supporting macro-cell deployments, and (3) GPUs for throughput-intensive High-PHY functions such as channel coding and decoding in emerging O-RAN platforms. While GPU-based acceleration is increasingly evaluated in O-RAN deployments, CPU-centric designs remain common in near-term vRAN systems due to energy efficiency and integration considerations.

In this broader vision, SIMD is not only an optimization but a cornerstone of scalable, agile, and maintainable PHY-layer processing. Its inherent compatibility with COTS servers and software stacks provides a pathway to realizing open, flexible, and high-performance RANs for 5G-Advanced and beyond. We hope this work encourages deeper exploration of SIMD-based acceleration strategies within the wireless research and communication engineering communities.

\section*{Acknowledgement}
This material is based upon work supported in part by the National Science Foundation under Grant No. NSF ECCS-2414678, in part by the Army Research Office under Grant W911NF2410107, and in part by the Korean government (RS-2024-00428780, RS-2024-00404972).

\bibliographystyle{IEEEtran}
\bibliography{jp_refs_mag}

\begin{IEEEbiographynophoto}{}
\noindent
\textbf{Jaebum Park} (Student Member, IEEE) received the B.S. degree from Yonsei University, Seoul, South Korea in 2015 and the M.S. degree from the Korea Advanced Institute of Science and Technology (KAIST), Daejeon, South Korea in 2017, both in Electrical and Electronic Engineering. From 2017 to 2024, he was with the 6G research team at Samsung Research, Seoul, South Korea. He is currently pursuing the Ph.D. degree with the Department of Electrical and Computer Engineering, University of California at San Diego, CA, USA. His research interests include advanced MIMO transceiver algorithms, energy-efficient MIMO techniques for next-generation RAN architectures, and digital predistortion.
\end{IEEEbiographynophoto}

\begin{IEEEbiographynophoto}{} 
\noindent
\textbf{Chan-Byoung Chae} (Fellow, IEEE) is an Underwood Distinguished Professor and Lee Youn Jae Fellow (Endowed Chair Professor) with the School of Integrated Technology at Yonsei University, Seoul, South Korea. He is also the Chief Scientific Officer (CSO) of SensorView, Ltd. He is the author of \emph{Signal Processing Engineering: An Intuitive Approach} (Springer, 2026) and has co-authored more than 200 journal papers. He has received several awards from CES, IEEE Communications Society (ComSoc), IEEE Signal Processing Society (SPS), and IEEE Vehicular Technology Society (VTS). He is an elected member of the National Academy of Engineering of Korea.
\end{IEEEbiographynophoto}

\begin{IEEEbiographynophoto}{} \noindent
\textbf{Robert W. Heath, Jr.} (Fellow, IEEE) is  the Charles Lee Powell Chair of Wireless Communications with the Department of Electrical and Computer Engineering, University of California at San Diego, CA, USA. He is also the President and the CEO of MIMO Wireless Inc. He has authored \emph{Introduction to Wireless Digital Communication} (Prentice Hall, 2017) and \emph{Digital Wireless Communication: Physical Layer Exploration Lab Using the NI USRP} (National Technology and Science Press, 2012) and coauthored \emph{Millimeter Wave Wireless Communications} (Prentice Hall, 2014) and \emph{Foundations of MIMO Communication} (Cambridge University Press, 2018).  He received the 2025 IEEE/RSE James Clerk Maxwell Medal. He is an elected member of the National Academy of Engineering in the USA.
\end{IEEEbiographynophoto}

\end{document}